\documentclass[12pt,a4paper]{book}
\usepackage{npcs}
\usepackage{graphics}
\usepackage{epsfig}

\ntitle{Non-zero Fermion Modes Contribution to the
Instanton-Induced Electroweak Processes at High Energies}

\nauthor{R.G.Shulyakovsky}

\naddress{B.I. Stepanov Institute of Physics,\\
 National Academy of Sciences of Belarus \\
\it F. Skorina avenue 68,\\
 220072 Minsk, Belarus \\
{\small \rm
   E-mails:     shul@dragon.bas-net.by}}

\ndata{
 }

\nabstract{We consider instanton-induced high energy processes in
an electroweak-type model. Multi-fermion final states are taken
into account in Gauss approximation. Unlike previous results we
calculate nonzero fermion modes contribution to cross-section by
straightforward procedure. This contribution can be important at
energies of sphaleron mass order.}

\nkeywords{ instantons, non-zero fermion modes, baryon number
violation}

\nPACSnumbers{ 11.15.Tk }
\begin{document}
\DeclareGraphicsExtensions{.jpg,.pdf,.mps,.png} 
\firstpage{2}  \nyear{2001}
\def\nfpage{\thepage}
\thispagestyle{myheadings} \npcstitle
\begin{sloppypar}
\section{Introduction}

As is well known~\cite{Hooft} instanton-induced processes in
electroweak theory can violate baryon number and concern the
problem of matter and antimatter asymmetry of the
Universe~\cite{Zuccero}. At the beginning of preceding decade it
was shown~\cite{Ringwald,Porrati,Zakharov} that cross-section of
these processes exponentially increase with energy (see for
review~\cite{Mattis})
\begin{equation}\label{Holy}
\sigma^{inst}(\varepsilon ) \sim
e^{\frac{16\pi^2}{g^2}F(\varepsilon )},\quad F(\varepsilon
)=-1+\frac{9}{8}\varepsilon^{\frac{4}{3}}-
\frac{9}{16}\varepsilon^{2}+...,\quad
\varepsilon=\frac{\sqrt{s}}{E_{sp}},
\end{equation}
where $\sqrt{s}$ is c.m. energy of the process,
$E_{sp}\approx\frac{4\sqrt{6}\pi m_w}{g^2}\approx 16\ TeV$ -
energy of sphaleron~\cite{Manton}. The first term in "Holy Grail"
function $F(\varepsilon )$ is 't~ Hooft suppression
factor~\cite{Hooft}, the second corresponds to the multiple $W$-
and Higgs bosons production in Gauss
approximation~\cite{Porrati,Zakharov}, the third is a contribution
of tree boson graphs~\cite{Khoze,Diak91}. Boson
loops~\cite{Rubakov} and fermion contribution~\cite{Espinosa} are
accumulated in pre-exponent factor. One can naively suppose that
$F(\varepsilon )\to 0$ if $\sqrt{s}\to E_{sp}$ and cross-section
(\ref{Holy}) can reach observable value~\footnote{Exact behavior
of "Holy Grail" is unknown at $\varepsilon \geq 1$ and a question
about possibility of observation of anomalous baryon number
violation is still opened.}. In this case pre-exponent constant in
(\ref{Holy}) can be important.

Contribution of W- and Higgs bosons diagrams to electroweak
instanton processes was calculated in detail (see
f.e.~\cite{Ringwald}-\cite{Mattis},\cite{Khoze}-\cite{Rubakov})
whereas graphs with arbitrary number of fermion legs were rarely
taken into account. In the pioneer paper of
O.Espinosa~\cite{Espinosa} author found multi-fermion contribution
by means of coherent-state formalism. In this paper we are
checking the results of Ref.~\cite{Espinosa} by straightforward
procedure in Gauss approximation and also finding the distribution
on fermion multiplicity.

\section{Multi-fermion Instanton-Induced Green Functions}

Let us consider electroweak-type model in the limit of vanishing
Weinberg angle with the following Euclidean action (we will
present most formulas in Euclidean space; analytical continuation
to the Minkovsky space will be performed in the final step)
\begin{equation}\label{Action}
S[A,\psi,\bar{\psi}]=S_{YM}+S_f+S_H,\quad S_{YM}=\frac{1}{2}\int
dx\, Tr\left(F_{\mu\nu}F_{\mu\nu}\right),\end{equation}
$$
S_f=-\int dx\sum_{i=1}^{n_f} \bar{\psi}i\!\not\!\!D\psi, \quad
S_H=\int dx\left((D_{\mu}\phi
)^+D_{\nu}\phi+\lambda\left(\phi^+\phi-\frac{v^2}{2}\right)^2\right),
$$
$$
\not\!\!D=\left(\partial_{\mu}-igA_{\mu}\right)\sigma_{\mu},\quad
F_{\mu\nu}=\partial_{\mu}A_{\nu}-\partial_{\nu}A_{\mu}-ig[A_{\mu},A_{\nu}],\quad
A_{\mu}=A^a_{\mu}\frac{\sigma_a}{2},
$$
where $n_f$ is a number of massless Weyl fermion doublets,
$n_f=12$ for three generations.

Let us consider for definiteness instanton-induced fermion
collisions with arbitrary number of fermions, $W$- and Higgs
bosons in final states
\begin{equation}\label{process}
f+f\to (2m-2)f+n_wW+n_hH.
\end{equation}
General form of Euclidean Green function for these
processes~\footnote{Only one sort of fermions is taken into
account. Generalization into $n_f=12$ is trivial.} can be given by
functional integral on gauge fields with one unit of topological
charge
\begin{equation}\label{Green}
\int\!DA D\phi D\psi D\bar{\psi}e^{-S}\prod_{i=1}^m\psi
(x_i)\prod_{j=m+1}^{2m}\bar{\psi}(x_j)\prod_{k=1}^{n_w}A_{\mu_k}(y_k)\prod_{l=1}^{n_h}\phi
(z_l),
\end{equation}
where group and spinor indexes are omitted for simplicity.

As is well known in Gauss approximation integration in
(\ref{Green}) on boson and fermion fields can be performed
independently
$$
\int d^4z\int_0^{\infty}\frac{d\rho}{\rho^5}\int
dU\mu(\rho)e^{-\frac{8\pi^2}{g^2}-\pi^2\rho^2v^2}
\prod_{k=1}^{n_w}A_{\mu_k}^{inst}(y_k-z)\prod_{l=1}^{n_h}\phi^{inst}
(z_l-z)\times
$$
\begin{equation}\label{Gauss}
\times\int D\psi D\bar{\psi}e^{\int dx\bar{\psi}i\not
D[A^{inst}]\psi}\prod_{i=1}^m\psi
(x_i-z)\prod_{j=m+1}^{2m}\bar{\psi}(x_j-z),
\end{equation}
where $z_{\mu},\, \rho,\, U$ and $\mu(\rho)$ are instanton center,
size, orientation and density correspondingly~\cite{Hooft, UFN},
$A^{inst}$ and $\phi^{inst}$ are instanton solutions~\cite{BPST}.

Since boson part of (\ref{Gauss}) was investigated quite well let
us concentrate on fermion contribution
\begin{equation}\label{fermion}
<\psi(x_1)...\bar{\psi}(x_{2m})>^{inst}\equiv \int D\psi
D\bar{\psi}e^{\int dx\bar{\psi}i\not
D[A^{inst}]\psi}\prod_{i=1}^m\psi
(x_i)\prod_{j=m+1}^{2m}\bar{\psi}(x_j),
\end{equation}
which can be considered as a fermion Green function in {\it
external} instanton field.

The simplest non-vanishing Green function for only one fermion
doublet reads~\cite{Hooft}
$$
<\psi(x_1)\bar{\psi}(x_{2})>^{inst}=\prod_n\int dc_nd\bar{c}_n
e^{\lambda_nc_n\bar{c}_n}
\sum_sc_s\psi_s(x_1)\sum_r\bar{c}_r\bar{\psi}_r(x_2)=
$$
\begin{equation}\label{1pair}
=det'(\rho)\psi_0(x_1)\bar{\psi}_0(x_2), \quad
det'(\rho)=\prod_{n\not=0}\lambda_n,\quad
i\not\!\!D[A^{inst}]\psi_n(x)=\lambda_n\psi_n(x),
\end{equation}
where Grassman algebra is used. Explicit expressions for zero mode
$\psi_0(x)$ (eigen function belonging to $\lambda_0=0$) and
$det'(\rho)$ are known~\cite{Hooft}.

Let us calculate 4-fermion Green function in the instanton field

$$
<\psi(x_1)\bar{\psi}(x_2)\psi(x_3)\bar{\psi}(x_4)>^{inst}=
\prod_n\int dc_nd\bar{c}_n e^{\lambda_nc_n\bar{c}_n}
\psi(x_1)\bar{\psi}(x_2)\psi(x_3)\bar{\psi}(x_4)=
$$
$$
=\int
dc_0d\bar{c}_0c_0\psi_0(x_1)\bar{c}_0\bar{\psi}_0(x_2)\prod_{n\not=0}\int
dc_nd\bar{c}_n(1+\lambda_nc_n\bar{c}_n)\psi(x_3)\bar{\psi}(x_4),
$$
$$
\prod_{n\not=0}\int
dc_nd\bar{c}_n(1+\lambda_nc_n\bar{c}_n)\psi(x_3)\bar{\psi}(x_4)=\sum_{k\not=0}\biggl(
\int dc_kd\bar{c}_kc_k\psi_k(x_3)\bar{c}_k\bar{\psi}_k(x_4)\times
$$
\begin{equation}\label{2pair}
\times\prod_{n\not=0,k}\int
dc_nd\bar{c}_n\lambda_nc_n\bar{c}_n\biggr)=\sum_{k\not=0}\Bigl(\psi_k(x_3)\bar{\psi}_k(x_4)\!\!\!
\prod_{n\not=0,k}\!\!\lambda_n \Bigr)=S^{nz}(x_3,x_4)det'(\rho),
\end{equation}
where non-zero fermion modes propagator $S^{nz}(x,y)$ can be
calculated analytically~\cite{BCCL,Shuryak}:
$$
S^{nz}(x,y)\equiv\sum_{j\not=0}\frac{\psi_j(x)\bar{\psi}_j(y)}{\lambda_j}=
\frac{1}{\sqrt{1+\rho^2/x^2}}\frac{1}{\sqrt{1+\rho^2/y^2}}\Biggl[\frac{(x-y)_{\mu}\sigma_{\mu}}{2\pi^2(x-y)^4}\times
$$
$$
\times\biggl(1+\rho^2\frac{x_{\nu}\sigma_{\nu}y_{\kappa}\bar{\sigma}_{\kappa}}{x^2y^2}
\biggr)-\frac{\rho^2}{4\pi^2(x-y)^2x^2y^2}\times
$$
\begin{equation}\label{nonzero}
\times\biggl(\bar{\sigma}_{\mu}\frac{x_{\nu}\sigma_{\nu}\bar{\sigma}_{\mu}(x-y)_{\lambda}\sigma_{\lambda}y_{\omega}
\bar{\sigma}_{\omega}}{\rho^2+x^2}+\sigma_{\mu}
\frac{x_{\nu}\sigma_{\nu}(x-y)_{\lambda}\bar{\sigma}_{\lambda}\sigma_{\mu}y_{\omega}
\bar{\sigma}_{\omega}}{\rho^2+y^2}\biggr)\Biggr],
\end{equation}
we use standard notations in Euclidean space
$\sigma_{\mu}=(-i\sigma_a,I),\, \bar{\sigma}_{\mu}=(i\sigma_a,I)$.
The propagator can be generalized to arbitrary instanton position
and color orientation.

 Finally,  for the 4-fermion Green function we obtain
\begin{equation}
<\psi(x_1)\bar{\psi}(x_2)\psi(x_3)\bar{\psi}(x_4)>^{inst}=\psi_0(x_1)\bar{\psi}_0(x_2)S^{nz}(x_3,x_4)
det'(\rho).
\end{equation}

Higher Green functions are calculated similarly:
$$
<\psi(x_1)\bar{\psi}(x_2)...\psi(x_{2m-1})\bar{\psi}(x_{2m})>^{inst}=
$$
\begin{equation}
=\psi_0(x_1)\bar{\psi}_0(x_2)S^{nz}(x_3,x_4)...S^{nz}(x_{2m-1},x_{2m})det'(\rho).
\end{equation}

\section{Multi-Fermion Contribution to the Observable Values}
Since boson part of the amplitude of the process (\ref{process})
is well known in Gauss approximation let us concentrate on the
fermion contribution only and consider for shortening that there
are not gauge and Higgs bosons in final state
\begin{equation}\label{process2}
f+f\to (2m-2)f.
\end{equation}
Generalization into arbitrary number of boson is trivial. For
calculation of the amplitude we can apply standard
Lehman-Symanzik-Zimmermann technique:
$$
A(k_1,p_1,...,k_m,p_m)=\int\limits_0^{\infty}\!\frac{d\rho}{\rho^5}\mu
(\rho) e^{-\frac{8\pi^2}{g^2}-\pi^2\rho^2v^2}
\tilde{\psi}_0(k_1)\tilde{\bar{\psi}}_0(p_1) \times
$$
\begin{equation}\label{Ampl}
\times
k_{1\eta}\sigma_{\eta}p_{1\eta'}\bar{\sigma}_{\eta'}det'(\rho)\prod_{j=2}^m
\tilde{S}^{nz}(p_j,k_j)k_{j\mu}\sigma_{\mu}p_{j\nu}\bar{\sigma}_{\nu}
\delta^{(4)}\Bigl(p_1+k_1-\sum_{n=2}^{m}(p_n+k_n)\Bigr),
\end{equation}
where $k_1$ and $p_1$ denotes 4-momenta of initial particles,
$k_2,p_2,...,k_m,p_m$ are 4-momenta of final fermions, tildes mean
Fourier transformations of corresponding functions,
$\delta$-function appears due to the integration over instanton
size $z_{\mu}$. It is supposed that all momenta satisfy the
mass-shell equation.

The contribution of the initial bosons and zero modes are well
known. Moreover, as is shown below, the leading fermion
contribution into cross-section comes from non-zero modes.
Therefore we restrict ourselves to non-zero modes contribution
into the amplitude (\ref{Ampl}). First of all Fourier
transformation of non-zero modes propagator (\ref{nonzero}) have
to be done
\begin{equation}\label{Fourier}\tilde{S}^{nz}(p,k)\equiv\int dxdye^{-ipx-ipy}S^{nz}(x,y)
=-2\pi^2\rho^2(p_{\mu}\bar{\sigma}_{\mu}+k_{\nu}\sigma_{\nu})
\frac{p_{\lambda}\sigma_{\lambda}k_{\lambda'}\bar{\sigma}_{\lambda'}}{p^2k^2(pk)},
\end{equation}
where we expand the expression in powers of $\rho^2$ and neglect
terms of order $O(\rho^4)$~\cite{Espinosa}. Amputation of external
legs and going to mass-shall gives
\begin{equation}\label{Amputed}
\lim_{p^2,\ k^2\to
0}\hat{S}^{nz}(p,k)k_{\eta}\sigma_{\eta}p_{\eta'}\bar{\sigma}_{\eta'}=
-2\pi^2\rho^2\frac{p_{\mu}\bar{\sigma}_{\mu}+k_{\nu}\sigma_{\nu}
}{(pk)}.
\end{equation}
Substituting (\ref{Amputed}) into (\ref{Ampl}) and performing
integration over $\rho$ we obtain
$$
A(k_1,p_1,...,k_m,p_m)=C(p_1,k_1)\ (m-1)!\times
$$
\begin{equation}\label{Ampl2}\times
\prod_{j=2}^m\left(\frac{-2}{v^2}\right)
\frac{p_{j\mu}\bar{\sigma}_{\mu}+k_{j\nu}\sigma_{\nu}}{(p_jk_j)}
\delta^{(4)}\Bigl(p_1+k_1-\sum_{n=2}^{m}(p_n+k_n)\Bigr),
\end{equation}
where zero modes are contained in factor $C$. Squaring and
integrating over 3-momenta of final particles we obtain the
probability of $m$ fermion pairs production in instanton field
$$
P_m=\int\limits_0^{\sqrt{s}}\frac{d\vec{p_1}}{(2\pi)^3}\frac{d\vec{k_1}}{(2\pi)^3}|C(p_1,k_1)|^2
(m-1)!\prod_{j=2}^m\frac{4}{v^4}
\int\limits_0^{\sqrt{s}}\frac{d\vec{p_j}}{(2\pi)^3}\frac{d\vec{k_j}}{(2\pi)^3}(p_jk_j)^{-1}\times
$$
$$
\times \delta^{(4)}\Bigl(p_1+k_1-\sum_{n=2}^{m}(p_n+k_n)\Bigr)=
\frac{N}{(m-1)!}\left(\frac{4}{v^4}4\pi\left(\frac{1-ln2}{3(2\pi)^6}\right)\frac{\sqrt{s}^6}{m_0^2}\right)^{m-1}=
$$
\begin{equation}\label{Prob}
=\frac{N}{(m-1)!}\left(\frac{\sqrt{s}^6}{m_0^2m_w^2E_{sp}^2}\frac{6(1-ln2)}{\pi^3}
\right)^{m-1},
\end{equation}
where small mass parameter $m_0$ is introduced for regularization
of the collinear singularity, condition $m_0\ll\sqrt{s}$ is used,
factor $N$ guarantees the probability normalization
$\sum\limits_{m=1}^{\infty}P_m=1$:
\begin{equation}\label{Norm}
N=\exp\left[-\frac{\sqrt{s}^6}{m_0^2m_w^2E_{sp}^2}\frac{6(1-ln2)}{\pi^3}\right].
\end{equation}

The contribution of fermions into total instanton-induced
cross-section with exponential accuracy is
\begin{equation}\label{cross}
\sigma^{inst}_f\sim\exp\left[\frac{n_f\sqrt{s}^6}{m_0^2m_w^2E_{sp}^2}\frac{6(1-ln2)}{\pi^3}
\right],
\end{equation}
where we took into account arbitrary number of fermions sorts.
Fig.1 shows the dependence of $ln\ \sigma^{inst}_f$ on energy.

\begin{minipage}{16.5cm}
\begin{center}
\epsfxsize=5in \epsfysize=3in \epsfbox{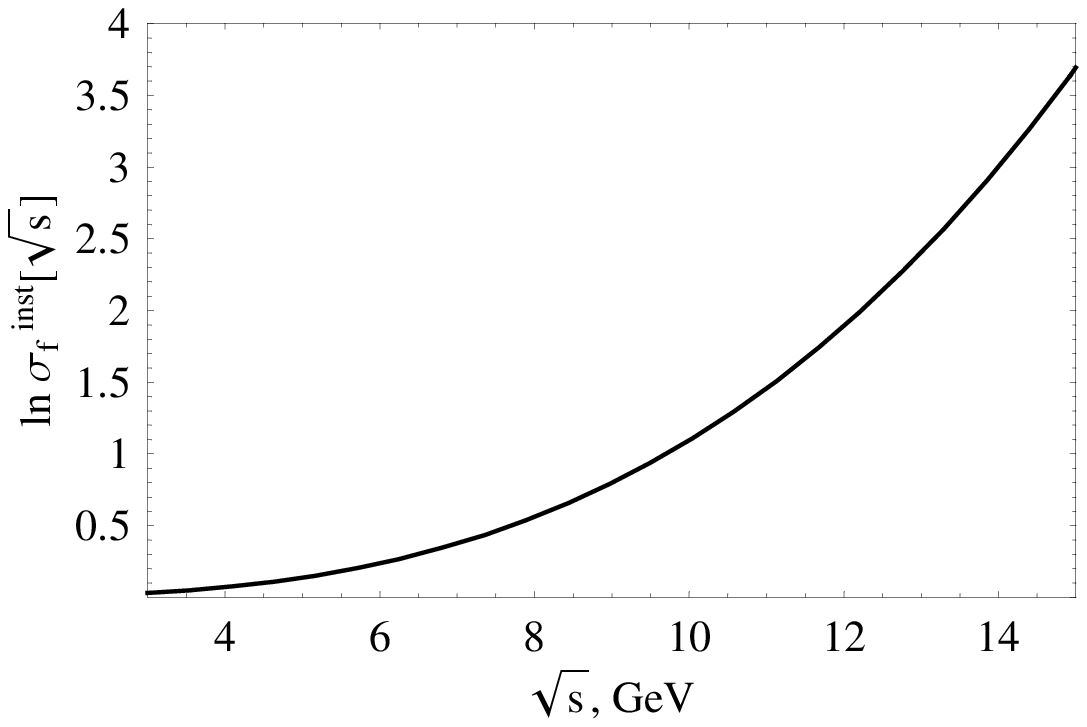} \end{center}
Fig.1. Logarithm of the fermion part of the total
instanton-induced cross-section in dependence on energy. We took
for definiteness $m_w=80\ GeV,\ m_0=0.5\ MeV,\ n_f=12$.
\end{minipage}

\section{Conclusion}
We obtain Poisson distribution on number of fermion pairs
(\ref{Prob}) produced in the instanton processes. Average number
of such pairs reads
\begin{equation}\label{average}
<m>=\frac{n_f\sqrt{s}^6}{m_0^2m_w^2E_{sp}^2}\frac{6(1-ln2)}{\pi^3}.
\end{equation}

Fermion part of instanton cross section (\ref{cross}) strongly
(exponentially) increases with energy. The contribution of zero
modes is included in pre-exponent factor. Non-zero fermion modes
are accumulated in exponent index and play essential role. At low
energies contribution (\ref{cross}) is not important. At high
energies $\sqrt{s}\sim\sqrt[3]{m_wm_0E_{sp}}\sim O(10)\, GeV$
exponent index is equal approximately to unit (see Fig.1). Of
course, at such energies fermion contribution is not essential due
to suppressing 't~ Hooft factor $exp[-\frac{16\pi^2}{g^2}]\approx
10^{-169}$. But at energies $\sqrt{s}\sim E_{sp}$ one can suppose
that 't~ Hooft suppression disappears and then strongly rising
fermion exponent have to be taken into account.

Fermion contribution into instanton-induced cross-section
calculated by means of coherent-state formalism by
O.Espinosa~\cite{Espinosa} also exponentially increase with energy
$\sigma_f^{inst}\sim~exp[\sqrt{s}^{4/3}]$. In comparison with the
result of Ref.~\cite{Espinosa} straightforward method gives
$(\sqrt{s})^6$ in exponent index. The difference can be caused the
approximations inherent to coherent-state method for fermionic
degrees of freedom.

\end{sloppypar}
\end{document}